\documentclass{llncs}
\usepackage{lineno}

\usepackage{simplebnf}	
\usepackage{listings}
\usepackage{xcolor,cmll}
\usepackage{paralist}
\usepackage{stmaryrd}
\usepackage{xspace,url,hyperref}
\usepackage{cite,xspace}
\usepackage{amsfonts,amssymb}
\usepackage{threeparttable,breakurl}
\usepackage[english]{babel} 
\usepackage{proof}

\usepackage{graphicx}
\usepackage{pgfplots}
\usepackage{lscape}
\usepackage{rotating}
\usepackage{epstopdf}

\newcommand{\instan}[1]{\hbox{\sl grnd\/}(#1)}
\newcommand{\Pscr}{{\mathcal P}}

\newcommand{\hastype}{\mathrel{:}}

\newcommand{\trueExpert}[1]{{\true_e}(#1)}
\newcommand{\eraseExpert }[1]{{\top_e}(#1)}

\newcommand{\unfoldExpert}[4]{{\hbox{\sl unfold\/}_e}(#1,#2,#3,#4)}
\newcommand{\andExpert}[3]{{\with_e}(#1,#2,#3)}
\newcommand{\tensorExpert}[3]{{\otimes_e}(#1,#2,#3)}

\newcommand{\bangExpert}[2]{\ensuremath{\bang_e(#1,#2)}}
\newcommand{\initExpert}[1]{\hbox{\textit{init}}_e(#1)}
\newcommand{\initExpertB}[1]{\hbox{\textit{init!}}_e(#1)}
\newcommand{\lolliExpert}[2]{\lolli_e\!(#1,#2)}

\newcommand{\true}{\mathbf{1}}
\newcommand{\XXi}{{\color{blue}{\Xi}}}

\newcommand{\passed}{{pass}}
\newcommand{\failed}{\textcolor{red}{found}}

\newcommand{\seq}[2]{#1\vdash #2}

\newcommand{\eval}[2]{#1 \Downarrow #2}
\newcommand{\ceval}[3]{(#1,#2) \Downarrow #3}
\newcommand{\cstep}[4]{(#1,#2) \leadsto (#3,#4)}
\newcommand{\has}[2]{#1 :   #2}
\newcommand{\hasv}[2]{#1\,  \textbf{:}\,   #2}
\newcommand{\rep}[1]{\ulcorner #1\urcorner}  
\newcommand{\state}{\sigma}

\def\one\mathit{true}
\def\di{\Delta_I}
\def\bnfas{\mathrel{::=}}

\def\arrow{\rightarrow}

\def\hastype{\mathrel{:}}
\def\lolli{\multimap}
\def\bang{\oc}
\newcommand{\io}[5]{#1 : #2\setminus #3 \vdash_{#4} #5}

\newcommand{\pair}[2]{\langle #1,#2\rangle}
                         {\endgroup]\end{quote}}

\long\def\ignore#1{}

\colorlet{lolli}{blue!70!black}
\definecolor{bleu}{HTML}{000DB3}
\lstdefinelanguage{lolli}{%
	alsoletter={-},
	classoffset=0,%
	morekeywords={},%
	keywordstyle=\color{lolli},%
	classoffset=0,%
	otherkeywords={<-,\&, x, ->,erase,true,bang},%
	sensitive=true,%
	morestring=[bd]",%
	morecomment=[l]\%,%
	morecomment=[n]{/*}{*/},%
}

\begin{document}
\title{Towards Substructural Property-Based Testing}
\author{Marco Mantovani \and Alberto Momigliano}
\institute{Dipartimento di Informatica,
  Universit\`{a} degli Studi di Milano, Italy
}
\maketitle
\begin{abstract}
  We propose to extend property-based testing to substructural logics
  to overcome the current lack of reasoning tools in the field. We
  take the first step by implementing a property-based testing system
  for specifications written in the linear logic programming language
  Lolli. We employ the foundational proof certificates architecture to
  model various data generation strategies. We validate our approach
  by encoding a model of a simple imperative programming language and
  its compilation and by testing its meta-theory via mutation
  analysis.
\end{abstract}
\section{Introduction}
\label{sec:intro}

Since their inception in the late 80's, logical frameworks based on
intuitionistic logic~\cite{fp-lf} have been successfully used to
represent/animate deductive systems ($\mathit{\lambda Prolog}$) and
also to reason (\emph{Twelf, Isabelle}) about them. The methodology of
\emph{higher-order abstract syntax} (HOAS) together with
parametric-hypothetical judgments~\cite{miller12proghol} yields
elegant encodings that lead to elegant proofs, since it delegates to
the meta-logic the handling of many common notions, in particular the
representation of \emph{contexts}. For example, when modeling a typing
system, we represent the typing context as a set of parametric
(atomic) assumptions: this tends to simplify the meta-theory since
properties such as weakening and context substitution come for free,
as they are inherited from the logical framework, and do not need to
be proved on a case-by-case basis. For an early example, see the
proof of subject reduction for MiniML in~\cite{MichaylovP91}, which
completely avoids the need to establish intermediate lemmas, as
opposed to more standard and labor-intensive
treatments~\cite{DBLP:conf/tphol/Dubois00}.

However, this identification of meta and object level contexts turns
out to be problematic in \emph{state-passing} specifications. To fix
ideas, consider specifying the operational semantics of an imperative
programming language: evaluating an  assignment requires taking an input
state, modifying it and finally returning it. A state (and related notions such as
heaps, stacks, etc.)  cannot be adequately encoded as a set of
intuitionistic assumptions, since it is intrinsically ephemeral. The
standard solution of reifing the state into a data structure, while
doable, betrays the whole HOAS approach.

Luckily, linear logic (and its substructural cousins) can change the
world, and in particular it provides a notion of context which has an
immediate reading in terms of resources.  A state \emph{can} be seen
as a set of linear assumptions and the linear connectives can be used
to model in a declarative way reading/writing said state.  In the
early 90's this idea was taken up in linear logic programming and
linear specification languages, viz., \emph{Lolli}~\cite{Hodas1994},
\emph{LLF}~\cite{CervesatoP02} and \emph{Forum}~\cite{miller96tcs}.

In the following years, given the richness of linear logic and
the flexibility of the proof-theoretic
foundations of logic programming~\cite{miller91apal}, more sophisticated languages
were proposed, with additional features such as order
(\emph{Olli}~\cite{Polakow00}), subexponentials~\cite{NigamM09},
bottom-up evaluation and concurrency
(\emph{Lollimon}~\cite{LopezPPW05},
\emph{Celf}~\cite{Schack-NielsenS08}).  Each extension required significant
 ingenuity, since it relied on appropriate notions of canonical
forms, resource management, unification etc. At the same time  tools for \emph{reasoning} over such substructural specifications
did not materialize, as opposed to the development of dedicated intuitionistic proof assistants such as
\emph{Abella}~\cite{BaeldeCGMNTW14} and
\emph{Beluga}~\cite{PientkaD10}.  
Meta-reasoning over such frameworks, in fact, asks for formulating
appropriate meta-logical tools, which, again, is far from
trivial. Sch\"{u}rmann et al.~\cite{McCreightS08} have designed a
linear meta-logics and Pientka et al.~\cite{GeorgesMOP17} have
introduced a linear version of contextual modal type theory to be used
within Beluga, but no implementations have appeared. The case for the
concurrent logical framework is particularly striking, where,
notwithstanding the wide range of applications~\cite{CLF2}, the only
meta-theoretic analysis available in \emph{Celf} is checking that a
program is well-moded.

If verification is too hard, or just while we wait for the field to
catch up, this paper suggests \emph{validation} as a useful
alternative, in particular in the form of \emph{property-based
  testing}~\cite{quickcheckfunprofit} (PBT). This is a lightweight validation
technique whereby the user specifies executable properties that the
code should satisfy and the system tries to refute them via automatic
(typically random) data generation.

Previous work~\cite{Blanco0M19} gives a proof-theoretic reconstruction
of PBT in terms of focusing and Foundational Proof Certificates
(FPC)~\cite{chihani17jar}, which, in theory applies to all the
languages mentioned above. The promise of the approach is that we can
state and check properties in the very logic where we specify them,
without resorting to a further meta-logic. Of course, validation is no
verification, but as by now common in mainstream proof assistants,
e.g.,~\cite{Bulwahn12,QChick}, we may
resort to testing not only \emph{in lieu of} proving, but
\emph{before} proving. 

In fact, the two-level architecture~\cite{gacek12jar} underlying the
\emph{Abella} system and the \emph{Hybrid} library~\cite{FeltyM12}
seems a good match for the combination of testing and proving over
substructural specifications. The approach keeps the meta-logic fixed,
while making substructural the specification logic. Indeed, some case
studies have been already carried out, the more extensive being the
verification of type soundness of quantum programming languages in a
Lolli-like specification logic~\cite{MahmoudF19}.

In this paper we move the first steps in this programme by
implementing PBT for {Lolli} and   evaluating its capability in
catching bugs by applying it to a mid-size case study: we give
a linear encoding of the static and dynamic semantic of an
imperative programming language and its compilation into a stack
machine and validate several properties, among which type
preservation and soundness of compilation. We have tried to  test properties in the way they would be stated and hopefully
proved in a linear proof assistant based on the two-level
architecture. That is, we are not arguing (yet) that linear PBT
is ``better'' than  traditional ones based on
state-passing specifications. Besides, in the case studies we
have carried out so far, we generate only \emph{persistent} data
(expressions, programs) under a given linear context.
Rather, we advocate the coupling of validation and (eventually)
verification for those encoding where linearity does make a difference
in terms of drastically simplifying the infrastructure one needs to
put in place towards proving the main result: one of the original
success stories of linear specifications, namely type preservation of
MiniMLR~\cite{mcdowell02tocl,CervesatoP02}, still stands and nicely
extends the cited one for MiniML: linearly, the theorem can be proven
from first principles, while with a standard encoding, for example the
Coq formalization in \emph{Software
  foundations}\footnote{\url{https://softwarefoundations.cis.upenn.edu/plf-current/References.html}},
you need literally dozens of preliminary lemmas.

In the following, we assume a passing familiarity with linear logic
programming and its proof-theory, as introduced in~\cite{Hodas1994,miller04llcs}.


\subsection{A motivating example}
\label{sec:mot}

\begin{figure}[t]
  \centering
    \[
         \infer[R_\to]{\Gamma\vdash A \to B}{\Gamma,A\vdash B}
      \qquad
         \infer[\mathit{init}]{\Gamma,A\vdash A}{}
    \]
    \[
      \infer[L^a_\to]{\Gamma,a\to B,a\vdash C}{\Gamma,B,a\vdash C}
      \qquad
      \infer[L^i_\to]{\Gamma,(A_1\to A_2) \to B\vdash C}{\Gamma, A_2\to B\vdash A_1\to A_2
        \quad \Gamma,B\vdash C}
    \]
          \dotfill
    \begin{lstlisting}
pv(imp(A,B)) <- (hyp(A) -> pv(B)).
pv(C) <- hyp(C) x erase.
pv(C) <- hyp(imp(A,B)) x atom A x hyp(A) x 
         (hyp(B) -> hyp(A) -> pv(C)).
pv(C) <- hyp(imp(imp(A1,A2),B)) x 
         (hyp(imp(A2,B) -> pv(imp(A1,A2))) & 
         (hyp(B) -> pv(C))).   
\end{lstlisting}

  \caption{Rules for contraction free LJF$_{\to}$ and their Lolli encoding}
  \label{fig:dyck}
\end{figure}

To preview our methodology, we present a self-contained example where
we use PBT as a form of \emph{model-based} testing: we evaluate an
implementation against a \emph{trusted} version.  We choose as trusted
model the linear encoding of the implicational fragment of the
\emph{contraction-free} calculus for propositional intuitionistic
logic, popularized by Dyckhoff. Figure~\ref{fig:dyck} lists the rules
for the judgment $\Gamma\vdash C$, together with a Lolli
implementation. Here, and in the following, we will use Lolli's
concrete syntax, where the arrow (in both directions) is linear
implication, \lstinline{x} is multiplicative conjunction (tensor),
\lstinline{&} is additive conjunction and \lstinline{erase} its unit
$\top$.

As shown originally in~\cite{Hodas1994}, we can encode provability
with a predicate \texttt{pv} that uses a linear context of
propositions \texttt{hyp} for assumptions at the left of the
turnstile, as shown in the first clause encoding the implication right
rule $R_\to$ via the embedded implication \lstinline|hyp(A) -> pv(B)|. In the left rules, the major premise is consumed by means of the tensor and the
new assumptions (re)asserted. Note that in rule $L^i_\to$, the context
$\Gamma$ is duplicated via additive conjunction. The \textit{init}
rule disposes via \lstinline|erase|of any remaining assumption since the object logic
enjoys weakening. By construction, the above code is a decision procedure for $\mathrm{LJF}_\to$.
%


Taking inspiration from Tarau's~\cite{Tarau19}, we consider next an
optimization where we factor the two left rule for implication in one:
\begin{lstlisting}
... % similar to before
pvb(C) <- hypb(imp(A,B)) x pvb_imp(A,B) x 
          (hypb(B) -> pvb(C)).
     
pvb_imp(imp(C,D),B) <- (hypb(imp(D,B)) -> pvb(imp(C,D))). 
pvb_imp(A,_)        <-  hypb(A).
\end{lstlisting}
\begin{sloppypar}
  Does the optimization preserve provability? Formally, the conjecture
  is
  \mbox{$\forall A \colon \mathit{form}.\ \mathit{pv(A)}\supset
    \mathit{pvb(A)}$}.  We could try to prove it, although, for the
  reasons alluded to in the introduction, it is not clear in which
  (formalized) meta-logic we would carry out such proof. Instead, it
  is simpler to test, that is to search for a counter-example. And the
  answer is no, the (encoding of the) optimization is faulty, as
  witnessed by the (pretty printed) counterexample
   \lstinline{A => ((A => (A => B)) => B)}: this intuitionistic tautology  fails to be provable in the purported
  optimization. We leave the fix to the reader.
\end{sloppypar}


\section{The proof-theory of PBT}
\label{sec:pt}

While PBT originated in  a functional programming
setting~\cite{claessen00icfp
}, at least two factors make a proof-theoretic reconstruction fruitful:
\begin{inparaenum}[1)]
\item it fits nicely with a (co)inductive reading of rule-based
  presentations of a system-under-test
\item it easily generalizes to richer logics.
\end{inparaenum}

If we view a property as a logical formula
\(\forall x [(\tau(x)\wedge P(x)) \supset Q(x)]\) where $\tau$ is a
typing predicate, 
providing a counter-example consists of negating the property, and
therefore searching for a proof of
\(\exists x [(\tau(x)\land P(x)) \land \neg Q(x)]\).

Stated in this way the problem points to a logic programming solution,
and since the seminal work of Miller et al.~\cite{miller91apal},
this means  proof-search in a focused sequent calculus, where the specification is a  set of
assumptions (typically sets of clauses) and the negated property is the query.

The connection of PBT with focused proof search is that in such a
query the \emph{positive phase} is represented by \(\exists x\) and
\((\tau(x)\land P(x))\). This corresponds to the generation of
possible counter-examples under precondition $P$.
That is followed by the \emph{negative phase} (which corresponds
to counter-example testing) and is represented by \(\neg Q(x)\). This
formalizes the intuition that generation may be arbitrarily hard, while testing is
just a deterministic computation.

How do we supply external information to the positive phase?  In
particular, how do we steer data generation? This is where the theory
of \emph{foundational proof certificates}~\cite{chihani17jar} (FPC)
comes in.  For the type-theoretically inclined, FPCs can be understood
as a generalization of proof-terms in the Curry-Howard
tradition. They have been introduced to  define and share a range of proof structures
 used in various theorem provers (e.g., resolution refutations,
 Herbrand disjuncts, tableaux, etc).  A FPC implementation consists of
 \begin{enumerate}
 \item    a generic proof-checking kernel,
 \item the specification of a certificate format, and  
 \item a set of predicates (called \emph{clerks and experts} to
   underline their different functionalities) that decorate the sequent
   rules used in the kernel and help to process the certificate.
 \end{enumerate}
In our setting, we can
   view those predicates as simple logic programs that guide the search for potential
counter-examples using different generation strategies.

 \subsection{Linear logic programming}


\begin{figure}[t]
  \centering
\[
  \infer{\io{\XXi}{\Delta_I}{\Delta_O}{} {G_1\with G_2}}
  {\io{\XXi_1}{\Delta_I}{\Delta_O}{} {G_1}\quad
    \io{\XXi_2}{\Delta_I}{\Delta_O}{} {G_2}\quad
    \andExpert{\XXi}{\XXi_1}{\XXi_2}} \qquad
  \infer{\io{\XXi}{\di}{\di}{} {\true}}
  {\trueExpert{\XXi}}
\]
\[
  \infer{\io{\XXi}{\Delta_I}{\Delta_O}{} {G_1\otimes G_2}}
  {\io{\XXi_1}{\Delta_I}{\Delta_M}{} {G_1}\quad
    \io{\XXi_2}{\Delta_M}{\Delta_O}{} {G_2}\quad
    \tensorExpert{\XXi}{\XXi_1}{\XXi_2}} \qquad
  \infer{\io{\XXi}{\di}{\Delta_O}{} {\top}}
{\Delta_I\supseteq\Delta_O\quad\eraseExpert{\XXi}}
\]
\[
  \infer{\io{\XXi}{\di}{\Delta_O}{}{A \lolli G}}
  {\io{\XXi'}{\di,A}{\Delta_O,\square }{}{G}\quad\lolliExpert{\XXi}{\XXi'}}
  \qquad
  \infer{\io{\XXi}{\di}{\di}{}{\bang G}}
    {\io{\XXi'}{\di}{\di}{}G \quad\bangExpert{\XXi}{\XXi'}}
  \]
    \[
      \infer{\io{\XXi}{\di,A}{\Delta_I,\square}{}A}
      {\initExpert{\XXi}}
      \qquad
        \infer{\io{\XXi}{\di,\bang A}{\Delta_O, \bang A}{}{A}}{\initExpertB{\XXi}}
      \]
\[
\infer{\io{\XXi}{\di}{\Delta_O}{} A}
      {\io{\XXi'}{\di}{\Delta_O}{} G 
        \quad \unfoldExpert{\XXi}{\XXi'}{A}{G}}
    \]


\caption{FPC presentation of the IO system for second order Lolli}
\label{fig:ce}
\end{figure}

Although focusing and FPC apply  to most sequent
calculi~\cite{LiangM09}, we find convenient to stay close to the
traditional semantics of uniform proofs~\cite{miller91apal}. The language that we adopt
here (shown below) is a minor restriction of linear Hereditary Harrop formul\ae\ which
underlay the linear logic programming language
{Lolli}~\cite{Hodas1994}. We consider implications with atomic
premises only and a first-order term language, thus making universal goals
essentially irrelevant.  The rationale of this is mirroring our Prolog
implementation, but we could easily account for the whole of Lolli.
\[
\begin{array}{llcl}
  \mbox{Goals}& G & \bnfas & A \mid \top \mid\true
                             \mid A \lolli G\mid \bang G\mid G_1 \otimes G_2\mid G_1 \with G_2  \\
  \mbox{Clauses}& D & \bnfas & \forall (G \arrow A)\\
  \mbox{Programs}& \Pscr & \bnfas & \cdot\mid \Pscr, D\\
      \mbox{Context}& \Delta & \bnfas & \cdot\mid \Delta, A\mid \Delta, \bang A\\
      \mbox{Atoms}& A & \bnfas & \dots
\end{array}
\]

This language induces an abstract logic programming language in the
sense of~\cite{miller91apal}, and as such can be given a uniform proof
system with a judgment of the form $\Delta \Rightarrow G$, for which
we refer once more to~\cite{Hodas1994}: intuitionistic implication
$A\rightarrow B$ is considered defined by $\bang A\lolli B$ and
therefore the intuitionistic context is omitted.

However, the uniform proofs discipline does not address the question
of how to perform proof search in the presence of linear assumptions,
a.k.a.\ the \emph{resource management
  problem}~\cite{CervesatoHP00}. The problem is caused by
multiplicative connectives that, under a goal-oriented strategy,
require a potentially exponential partitioning of the given linear
context.

One solution, based on \emph{lazy} context splitting and known as the
\emph{IO system}, was introduced in~\cite{Hodas1994}, and further
refined in~\cite{CervesatoHP00}: when we need to split a context (here
only in the tensor case), we give to one of the sub-goal the whole
input context ($\di$): some of it will be consumed and the leftovers
($\Delta_O$) returned to be used by the other sub-goal.

Figure~\ref{fig:ce} contains a version of the IO system for our
language as described by the judgment
$\io{\Xi}{\Delta_I}{\Delta_O}{}{G}$, where $\setminus$ is just a
suggestive notation to separate input and output context.  We will
explain the role of $\Xi$ and the predicates
$\mathit{op}_e( \Xi,\dots )$ in the next paragraphs. We overload
``$,$'' to denote multi-set union and adding a formula to a
context. Following on the literature and our
implementation, we will signal that a resource has been consumed in
the input context by replacing it with the placeholder ``$\square$''.

The IO system (without certificates) is known to be sound and complete
w.r.t.~uniform provability: $\di\setminus\Delta_O\vdash G$ iff
$\di - \Delta_O\Rightarrow G$, where ``$-$'' is a notion of context
difference modulo $\square$ (see~\cite{Hodas1994} for details). Given
this relationship, the requirement for the linear context to be empty
in the right rules for $\true$ and $\bang$ is realized by the notation
$\di\setminus\di$. In particular, in the linear axiom rule, $A$ is the
only available resource, while in the intuitionistic case, $\bang A$
is not consumed. The tensor rule showcases lazy context splitting,
while additive conjunction duplicates the linear context.

The handling of $\top$ is sub-optimal, since it succeeds with any
subset of the input context. As well known~\cite{CervesatoHP00}, this
could be addressed by moving to a system with \emph{slack}. However,
given the preferred style of our encodings (see
Section~\ref{sec:imp}), where additive unit is called only as a last
step, this has so far not proved necessary.

Building on the original system and in accord with the FPC approach,
each inference rule is augmented with an additional premise involving
an {expert predicate}, a certificate $\Xi$, and possibly resulting
certificates ($\Xi'$, $\Xi_1$, $\Xi_2$) reading the rules from
conclusion to premises. Operationally, the certificate $\Xi$ is an
input in the conclusion of a rule and the continuations are computed
by the expert to be handed over to the premises, if any.
%

The FPC methodology requires first to describe a format for the certificate.
Since we use FPC only to guide proof-search, we fix
the following three formats and we allow their composition, known as \emph{pairing}:
\[
\begin{array}{llcl}
  \mbox{Certificates}& \XXi & \bnfas & n \mid \langle n, m \rangle \mid d\mid (\XXi,\XXi)
\end{array}
\]
The first certificate is just a natural number and it used to bound a
derivation according to its \emph{height}. Similarly, the second
consists of a pair of naturals that bounds the number of clauses used
in a derivation (\emph{size}): typically $n$ will be input and $m$
output, so the \emph{size} will be $n-m$. In the third case, $d$ stands
for a \emph{distribution} of weights to clauses in a predicate
definition, to be used for random generation; if none is given, we
assume a uniform distribution. Crucially, we can compose
certificates, so that for example we can provide random generation bounded
by the height of the derivation; pairing is a simple, but surprisingly effective combinator~\cite{Pair}.

Each certificate format is accompanied by the implementation of the
experts that process the certificate in question.  
We exemplify the FPC discipline with a selection of  rules instantiated with the
\emph{size} certificates. If we run the judgment $\io{\langle n, m \rangle}{\di}{\Delta_O}{}{G}$,  the inputs are $n,\ \di$ and $G$, while $\Delta_O$ and $m$ will be output.

\[
\infer{\io{\pair n m}{\di}{\Delta_O}{} A}
      {\io{\pair {n - 1}{m}}{\di}{\Delta_O}{} G \quad (A\leftarrow G)\in\instan\Pscr
        \quad n > 0}
      \qquad
  \infer{\io{\pair n n}{\di}{\di}{} {\true}}
  {}
    \]
    \[
      \infer{\io{\pair i o}{\Delta_I}{\Delta_O}{} {G_1\otimes G_2}}
  {\io{\pair i m}{\Delta_I}{\Delta_M}{} {G_1}\quad
    \io{\pair m o}{\Delta_M}{\Delta_O}{} {G_2} }
\]
\[
    \infer{\io{\pair n m}{\Delta_I}{\Delta_O}{} {G_1\with G_2}}
  {\io{\pair n m}{\Delta_I}{\Delta_O}{} {G_1}\quad
    \io{\pair n m}{\Delta_I}{\Delta_O}{} {G_2}} 
  \]
  Here (as in all the formats considered in this paper), most experts are
  rather simple; they basically hand over the certificate according to
  the connective. This is the case of $\with$ and $\true$, where the
  expert copies the bound and its action is implicit in the
  instantiation of the certificates in the premises. In the tensor
  rule, the certificate mimics context splitting.
The \emph{unfold} expert, instead, is more interesting: not only does
it decrease the bound, provided we have not maxed out on the latter, but
it is also in charge of selecting the next goal: for bounded search via
chronological backtracking over the grounding of the program.  This
very expert is also the hook for implementing random data generation via
random back-chaining, where we replace chronological with randomized
backtracking: every time the derivation reaches an atom, we permute
its definition and pick a matching clause according to the
distribution described by the certificate. Other strategies are
possible, as suggested in~\cite{pltredexconstraintlogic}: for example,
permuting the definition just once at the beginning of generation, or
even randomizing the conjunctions in the body of a clause.

Note that we have elected \emph{not} to delegate to the experts
resource management: while possible, it would force us to pair such
certificate with any other one.
 As detailed in~\cite{Blanco0M19}, more sophisticated FPCs
capture other features of PBT, such as $\delta$-debugging (shrinking) and
bug-provenance, and will not be repeated here.

We are now ready to account for the soundness property from the example in
Section~\ref{sec:mot}. By analogy, this applies to  certificate-driven PBT with a
liner IO kernel in general. Let $\XXi$ be here the height certificate with bound $4$ and
\texttt{form} a unary predicate describing the syntax of implication
formul\ae, which we use as a generator. Testing the property  becomes the following  query in a
host language that implements the kernel:
  \[
    \exists F.\ (\io{\XXi}{\cdot}{\cdot}{}{\texttt{form(F)}})\land \, 
    (\io{\XXi}{\cdot}{\cdot}{}{\texttt{pv(F)}}) \land \, 
    \neg (\io{\XXi}{\cdot}{\cdot}{}{\texttt{pvb(F)}})
    \]
    In our case, the meta-language is simply Prolog, where we encode
    the kernel with a predicate \texttt{prove/4} and to
    check for un-provability
    negation-as-failure suffices,  as argued in~\cite{Blanco0M19}.
\begin{lstlisting}[language=prolog]
C = height(4),prove(C,[],[],form(F)),prove(C,[],[],pv(F)),
\+ prove(C,[],[],pvb(F)).
\end{lstlisting}
  


    \section{Case study}
\label{sec:imp}

IMP is a model of a minimalist Turing-complete imperative programming
language, featuring instructions for assignment, sequencing,
conditional and loop. It has been extensively used in teaching and in
mechanizations (viz.~formalized textbooks such as \emph{Software
  Foundations} and \emph{Concrete Semantics}). Here we follow Leroy's
account~\cite{Leroy10}, but add a basic type system to distinguish
arithmetical from Boolean expressions. IMP is a good candidate for a
linear logic encoding, since its operational semantics is, of course,
state based, while its syntax (see below) is simple enough not to
require a sophisticated treatment of binders.

\begin{bnfgrammar}
	expr ::= 
	  var											: variable
	| i										: integer constant
	| b                        : Boolean constant
	| expr $+$ expr 									: addition
	| expr $-$ expr										: subtraction
	| expr $*$ expr										: multiplication
	| expr $\wedge$ expr								: conjunction
	| expr $\vee$ expr									: disjunction	
	| $\neg$ expr										: negation
	| expr $==$ expr									: equality
\end{bnfgrammar}
\begin{minipage}[t]{0.5\linewidth}
\begin{bnfgrammar}
	val ::= 
	| vi										: integer value
	| vb										: Boolean value
\end{bnfgrammar}
\end{minipage}
\begin{minipage}[t]{0.6\linewidth}
\begin{bnfgrammar}
	ty ::= 
	| tint										: type of integers 
	|tbool										: type of Boolean's
\end{bnfgrammar}
\end{minipage}
\begin{bnfgrammar}
	cmd ::= 
	  skip				 								: no op
	| cmd ; cmd									: sequence
	| if expr then cmd else cmd			 		: conditional
	| while expr do cmd		 						: loop
	| var = expr 									: assignment
\end{bnfgrammar}	

The relevant judgments describing the dynamic and static semantics of IMP are:
\begin{description}
\item[$\seq\state\eval m v$] big step evaluation of expressions;
\item[$\ceval c {\state} \state'$] big step execution of commands;
\item[$\cstep c {\state} {c'} {\state'}$] small step execution of
  commands and its Kleene closure;
\item[$\seq\Gamma\has m \tau$]  well-typed expressions and   $\hasv v \tau$  well-typed values;
  \item[$\seq\Gamma c$] well-typed commands and  $\has\Gamma \state$ well-typed states;
\end{description}


\subsection{On linear encodings}
\label{ssec:enc}

In traditional accounts, a state $\state$ is a (finite) map
between variables and values. Linear logic  takes a
``distributed'' view and represent a state as a multi-set of linear
assumptions.  Since this is central to our approach, we make explicit
the (overloaded) encoding function $\rep\cdot$ on states. Its action
on expressions and values is as expected and therefore omitted:
\[
\begin{array}{rcl}
  \state  &\bnfas&  \cdot\mid \state, \ x \mapsto v\\
   \rep{\cdot} & = & \emptyset\\
      \rep{ \state, x \mapsto v} & = & \rep{\state},\mathit{var( x, \rep{v})}
  \end{array}
\]

When encoding state-based computations such as evaluation 
and execution 
in a Lolli-like language, it is almost forced on us to use a
\emph{continuation-passing style} (CPS): by sequencing the
computation, we get a handle on how to express ``what to compute
next'', and this  turns out to be the right
tool to encode the operational semantics of state update. 
CPS fixes a given evaluation order, which is crucial when the modeled
semantics has side-effects, lest adequacy is lost.

Yet, even under the CPS-umbrella, there are choices: e.g., 
whether to adopt an encoding that privileges \emph{additive}
connectives, in particular when using the state in a non-destructive way. In the additive
style, the state is duplicated with $\with$ and then eventually
disposed of via $\top$ at the leaves of the derivation.

This is well-understood, but, at least in our setup,
%
it leads to the
reification of the continuation as a data structure and the
introduction of an additional layer of instructions to manage the
continuation: for an example, see the static and dynamic semantics of
MiniMLR in~\cite{CervesatoP02}\footnote{This can be circumvented by
  switching to a more expressive logic, either by internalizing the
  continuation as an ordered context~\cite{Polakow00} or by changing
  representation via forward chaining (\emph{destination-passing
    style})~\cite{LopezPPW05}.}.

Mixing additive and multiplicative connectives needs a more sophisticated
resource management system; this is a concern, given
%
%
the efficiency requirements that testing brings to the table --- it is
not called ``QuickCheck'' for nothing. We therefore use the notion of
\emph{logical} continuation advocated by
Chirimar~\cite{chirimar95phd}, which affords us the luxury to never
duplicate the state. Logical continuations need higher-order logic (or
can be simulated in an un-typed setting such as Prolog).  Informally,
the idea is to transform every atom $A$ of type
$(\tau_1 * \dots * \tau_n)\rightarrow o$ into a new one $\hat A$ of
type $(\tau_1 * \dots * \tau_n* o) \rightarrow o$ where we accumulate
in the additional argument the body of the definition of $A$ as a
nested goal.
Facts are transformed so that
the continuation becomes the precondition.

For example, consider a fragment of the rules for the evaluation judgment
$\seq\state\eval m v$ and its encoding:

  \[
  \begin{array}{c}
    \infer[e/v]{\seq{\state}{ \eval x n}}{x \mapsto n\in\state}\qquad             \infer[e/n]{\seq{\state}{ \eval n n}}{} \smallskip\\
    \infer[e/p]{\seq{\state}{\eval{ e_1 + e_2} {v}}} 
    {\seq{\state}{\eval {e_1} v_1}\quad\seq{\state}{ \eval {e_2}{v_2}}\quad \mathrm{ plus\ v_1  \ v_2\ v}}
  \end{array}
\]
\begin{lstlisting}
eval(v(X),N,K)            <- var(X,N) x (var(X,N) -> K).
eval(i(N),vi(N),K)        <- K.
eval(plus(E1,E2),vi(V),K) <- 
   eval(E1,vi(V1),eval(E2,vi(V2),bang(sum(V1,V2,V,K)))).  
\end{lstlisting}
In the variable case, the value for \lstinline|X| is read (and
consumed) in the linear context and consequently reasserted; then we
call the continuation in the restored state. Evaluating a constant
\lstinline|i(N)| will have the side-effect of instantiating
\lstinline|N| in \lstinline|K|. The clause for addition showcases the
sequencing of goals inside the logical continuation, where the \lstinline|sum|
predicate is ``banged'' as a computation that does not need the state.

The \emph{adequacy} statement for CPS-evaluation reads:
$\seq{\state}{\eval m v}$ iff the sequent
$\rep{\state}\Rightarrow \mathtt{eval}(\rep{m},\rep{v}, \top)$ has a
uniform proof, where the initial continuation $\top$ cleans up
$\state$ upon success. As well-know, we need to generalize the
statement to arbitrary continuations for the proof to go through.

It is instructive to look at an additive encoding as well:
\begin{lstlisting}
ev(v(X),V)            <- var(X,V) x erase.
ev(i(N),vi(N))        <- erase.
ev(plus(E1,E2),vi(V)) <- ev(E1,vi(V1)) &
                         ev(E2,vi(V2)) &
                         bang(sum(V1,V2,V)).
\end{lstlisting}
While this seems appealingly simpler, it breaks down when the state is
updated and not just read;
consider the operational semantics of  {assignment} and its encoding:
\[
      \infer[]{\ceval{\state} {x := m } \state \oplus \{x \mapsto v\}}{\seq{\state}{\eval m v}}
\]
\begin{lstlisting}
ceval(asn(X,E),K) <- eval(E,V,(var(X,_) x (var(X,V) -> K))).
\end{lstlisting}
The continuation is in charge of both having something to compute
after the assignment returns, but also of sequencing in the right
order reading the state via evaluation, and updating via the embedded
implication. An additive encoding via $\with$ would not be adequate, since the connective's commutativity is at odd with side-effects.

At the top level, we initialize the execution of programs (seen as a
sequence of commands) by using as initial continuation  a predicate
\lstinline+collect+ that consumes the final state and
returns it  in a reified format.
\begin{lstlisting}
main(P,Vars,S) <- ceval(P,collect(Vars,S)).
\end{lstlisting}

We are now  in the position of addressing  the meta-theory
of our system-under-study via testing.  We list the more important properties
among those that we have considered. All statements are universally
quantified:
\begin{itemize}
  	\item[\textbf{srv}] subject reduction for evaluation:
		$
			\seq{\Gamma}{m \hastype \tau} \longrightarrow \seq{\state}{ \eval m v} \longrightarrow \has\Gamma \state \longrightarrow \hasv v \tau
		$; 

	\item[\textbf{dtx}] determinism of execution:
		$
		\ceval{\state} c  {\state_1} \longrightarrow \ceval {\state}c {  \state_2} \longrightarrow \state_1 \approx \state_2
		$;
              \item[\textbf{srx}] preservation of state under
                execution:
                $ \seq{\Gamma}{c} \longrightarrow \has\Gamma\state
                \longrightarrow \ceval{\state}c { \state'}
                \longrightarrow \has\Gamma\state' $;
	\item[\textbf{pr}] progress for small step execution:
		$
                \seq{\Gamma}{c} \longrightarrow \has\Gamma \state \longrightarrow c = \mathtt{skip} \vee \exists c'\, \state',\, (c,\state)\leadsto (c', \state')
		$;
	\item[\textbf{eq}] equivalence of small and big step execution (assuming determinism of both):
		$
			\ceval{\state}c { \state_1} \longrightarrow (c, \state_1) \leadsto^* (\mathtt{skip},\state_2) \longrightarrow \state_1 \approx \state_2
		$.
\end{itemize}

We have also encoded the compilation of IMP to a stack machine and
(mutation) tested forward and backward simulation of compilation
w.r.t.\ source and target execution. We have added a simple type
discipline for the assembly language in the spirit of
TAL~\cite{MorrisettWCG99} and tested preservation and progress, 
to exclude underflows in the execution of a well-typed
stack machine. Details can be found in the accompanying repository.



    \subsection{Experimental evaluation}
\label{ssec:exper}



 A word of caution before discussing our experiments: first, we have
spent almost no effort in crafting nor tuning custom generators; in fact,
they are simply FPC-driven regular unary logic
programs~\cite{YARDENI} with a very minor massage. Compare this with
the amount of ingenuity poured in writing generators in~\cite{TNIQ} or with
the model-checking techniques of~\cite{RobersonHDB08}. Secondly, our
interpreter is a Prolog meta-interpreter and while we have tried to
exploit Prolog's indexing, there are obvious ways to improve its efficiency, from
partial evaluation to better data structures for
contexts.

To establish a fair baseline, we have also implemented a
``vanilla'' version of our benchmarks, that is  \emph{state-passing}
ones, driven by a FPC-lead vanilla meta-interpreter.
%
%
%
We have run the experiments on a laptop with an Intel i7--7500U CPU and
16GB of RAM running WSL (Ubuntu 20.04) over Windows 10, using
SWI-Prolog 8.2.4.  All times are in seconds, as reported by SWI's
\texttt{time/1}. They are the average of five measurements.
We list here only a few experiments with no pretense of completeness. In
particular, we choose a fixed exhaustive generation strategy 
(\emph{size}), then pair it with \emph{height} for mutation analysis. We use
consistently certain bounds that experimentally have shown to be effective in
generating enough interesting data.

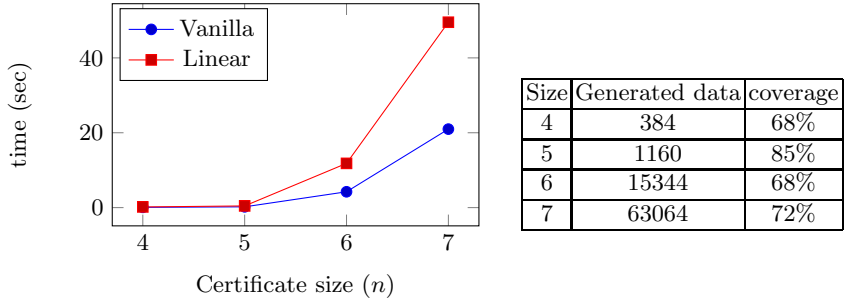
\begin{figure} [t]
	\centering
	\begin{minipage}{0.49\linewidth}\centering
		\begin{tikzpicture}
			\begin{axis}[
				xlabel={Certificate size ($n$)},
				ylabel={time (sec)},
				xtick={4,5,6,7},
				width=1.08\textwidth,
				height=0.759\textwidth,
				legend entries={Vanilla,Linear},
				legend style={at={(0.02,0.98)},
					anchor=north west}
				]
				\addplot table{
					4 0.07
					5 0.21
					6 4.23
					7 20.97
				};
				\addplot table{
					4 0.18
					5 0.45
					6 11.82
					7 49.53
				};
			\end{axis}
		\end{tikzpicture}
	\end{minipage}
	\begin{minipage}{0.49\linewidth}\centering
		\begin{tabular}{|c|c|c|}
			\hline	
			Size & Generated data & coverage \\
			\hline
			4 & 384 & 68\% \\
			\hline
			5 & 1160 & 85\% \\
			\hline	
			6 & 15344 & 68\% \\
			\hline	
			7 & 63064 & 72\% \\
			\hline	
		\end{tabular}

              \end{minipage}
              		\caption{Testing property \textbf{eq} with certificate $\langle n,\_\rangle$}
		\label{fig:tess}
\end{figure}

First we compare the time to test a sample property (``\textbf{eq}'',
the equivalence of big and small step execution) over a bug-free model
both with linear and vanilla PBT. On the left of Fig.~\ref{fig:tess}
we plot the time proportionally to the certificate size. On the right
we list the number of generated programs and the percentage of those
that converge within a bound given by a polynomial function over the
certificate size. The linear interpreter performs worse than the state
passing one, but not dramatically so. This is to be expected, since
the vanilla meta-interpreter does not do context management: in fact, it
does not use logical contexts at all.

\begin{figure}[t]
  \centering
    \begin{small}
\begin{tabular}{|c|c|c|c|c|c|c|}
	\hline
	& \textbf{dtx} & \textbf{srx} & \textbf{srv} & \textbf{pr} & \textbf{eq} & \textbf{cex} \\
	\hline
	M1 
		&
			\passed			
		&
			\passed
	 	&
	 		\passed
	 	&
	 		\passed
	 	&  
	 		\passed
	 	\\
	\hline
	M2
		&
			\failed		
		&
			\passed
		&
			\passed
		&
			\passed
		&  
           \failed
           & \lstinline!w := 0 - 1!
		\\
	\hline
	M3 
		&
			\passed
		&
			\passed
		&
			\passed
		&
			\passed
		&  
			\passed
		\\
	\hline
	M4 
		&
			\passed
		&
			\failed
		&
			\failed
		&
			\passed
		&  
           \passed
           & \lstinline!x := tt /\ tt! 
		\\
	\hline
	M5
		&
			\passed
		&
			\passed
		&
			\passed
		&
			\passed
		&  
			\passed
		\\
	\hline
	M6
		&
			\failed
		&
			\passed
		&
			\passed
		&
			\passed
		&  
           \failed
            & \lstinline!if x = x then {w := 0} else {w := 1}!
		\\
	\hline
	M7
		&
			\passed
		&
			\passed
		&
			\passed
		&
			\passed
		&  
			\passed
		\\
	\hline
	M8
		&
			\passed
		&
			\passed
		&
			\passed
		&
			\passed
		&  
			\passed
		\\
	\hline
	M9 
		&
			\passed
		&
			\passed
		&
			\passed
		&
			\passed
		&  
			\passed
		\\
	\hline
\end{tabular}
\end{small}

\caption{Mutation testing}
\label{fig:mut}
\end{figure}

Next, to gauge the effectiveness in catching bugs, we use, as customary,
\emph{mutation analysis}~\cite{Jia:2011}, whereby single intentional
mistakes are inserted into the system under study. A testing suite
is deemed as good as its capability of detecting those bugs (\emph{killing a
  mutant}). Most of the literature about mutation analysis revolves
around automatic mutant analysis for imperative code, certainly not
(linear) logical specifications of object logics. Therefore, we resort
to the \emph{manual} design of a small number of mutants, with all the limitations
entailed. Note, however, that this is the approach taken by the
testing
suite\footnote{\url{https://docs.racket-lang.org/redex/benchmark.html}}
of a comparable tool such as \emph{PLT-Redex}~\cite{PLTbook}. The
mutations are described in Appendix~\ref{app:m}.

Table~\ref{fig:mut} summarizes the outcome of mutation testing, where
``\failed'' indicates that a counter-example (cex) has been found and
``\passed'' that the bound has been exhausted. In the first case, we
report  counter-examples in the last column, after pretty-printing.  Since this is
accomplished in milliseconds, we omit the precise timing
information. Note that cex found by exhaustive search are minimal by
construction. 

The results
seem at first disappointing ($3$ mutants out of $9$  being detected), until we realize
that it is not so much a question of our tool failing to kill mutants,
but of the above properties being too loose. Consider for example
mutation M3: being a type-preserving operation swap in the evaluation of
expressions, this will certainly not lead to a failure of subject
reduction, nor invalidate determinism of evaluation.
On the other hand all mutants are easily killed with model-based
testing, that is taking as properties soundness and completeness of
the (top-level) judgments where mutations occur w.r.t.\ their bug-free
versions executed under the vanilla interpreter. This is reported in
Table~\ref{fig:mbt}.
\begin{figure}[ht]
  \renewcommand{\passed}{{pass }}
\renewcommand{\failed}{\textcolor{red}{found }}

  \centering
  \begin{center}
	\begin{tabular}{|c|c|c|c|}
		\hline
          & exec: $C \to L$ & exec: $L\to C$ & cex
          \\
		\hline
		No Mut 
		&
			\passed in 2.40
		&
           \passed in 6.56
		\\	
		\hline
		M1
		&
			\failed in 0.06
		& 
           \passed in 6.45
                      & \lstinline!w := 0 + 0!
		\\
		\hline
		M2
		&
			\passed in 2.40
		&
           \failed in 0.04
&            \lstinline!w := 0 - 1! 
		\\
		\hline
		M3
		&
			\failed in 0.06
		&
           \failed in 0.06
            & \lstinline!w := 0 * 1!
		\\
		\hline
		M4
		&
			\failed in 0.06
		&
           \failed in 0.04
           & \lstinline!y := tt /\ tt!
		\\
		\hline
		M5		
		&
			\failed in 0.00
		&
           \passed in 5.15
           & \lstinline!w := 0; w := 1!
		\\
		\hline
		M6
		&
			\passed in 2.34
		&
           \failed in 0.17
&           \lstinline!if y = y then {w := 0} else {w := 1}!
		\\
		\hline
		M7
		&
			\failed in 0.65
		&
           \passed in 0.82
           & \lstinline!while y = y /\ y = w do {y := tt}!
		\\
		\hline
	\end{tabular}

	\begin{tabular}{|c|c|c|c|}
		\hline
		& type: $C \to L$ &  type: $L\to C$ & cex\\
		\hline
		No Mut 
		&
			\passed in 0.89
		&
			\passed in 0.87
		\\	
		\hline
		M8
		&
			\failed in 0.03
		&
           \passed in 0.84
           & \lstinline!w := 0 + 0!
		\\
		\hline
		M9
		&
			\failed in 0.04
		&
           \passed in 0.71
           & \lstinline!y := tt \/ tt! 
		\\
		\hline
	\end{tabular}
\end{center}

\caption{Model-based testing of  IMP mutations}
\label{fig:mbt}
\end{figure}



\section{Conclusions}
\label{sec:what}

In this paper we have argued for the extension of property-based
testing to substructural logics to overcome the current lack of
reasoning tools in the field. We have taken the first step by
implementing a PBT system for specifications written in linear
Hereditary Harrop formul\ae, the language underlying Lolli. We
have adapted the FPC architecture to model various generation
strategies. We have validated our approach by encoding the meta-theory
of IMP and its compilation, with a rudimentary mutation analysis. With
all the caution that our setup entails, results are encouraging.

There is so much future work that it is almost overwhelming: first
item from the system point of view is abandoning the
meta-interpretation approach, and then a possible integration with
Abella. Theoretically, our plan is to extend our framework to richer
linear logic languages, featuring ordered logic up to
concurrency, as well as supporting different operational semantics, to
begin with bottom-up evaluation.

\smallskip
\noindent
\begin{sloppypar}
Source code can be found at
\burl{https://github.com/Tovy97/Towards-Substructural-Property-Based-Testing}
\end{sloppypar}
\paragraph{Acknowledgments}
We are grateful to Dale Miller for many discussions and in particular
for suggesting the use of logical continuations. Thanks also to Jeff
Polakow for his comments on a draft version of this paper.



\appendix

\section{Appendix}
\label{sec:app}


 \paragraph{List of mutations}
 \label{app:m}

We describe a selection of the mutations that we have implemented,
together with a categorization, borrowed from the classification of
mutations for Prolog-like languages in~\cite{MomiglianoO19}. We also
report the judgment where the mutation occurs.
\begin{description}
\item[{\underline{Cl}}ause mutations:]  deletion of a predicate in the body of a clause,
  deleting the whole clause if a fact. 
\item[\underline{O}perator mutations:] arithmetic and relational operator
  mutation. 
\item[\underline{V}ariable mutations:] replacing a variable with an (anonymous)
  variable and vice versa.
\item[\underline{C}onstant mutations:] replacing a constant by a
  constant (of the same type), or by an (anonymous) variable and vice
  versa.
\end{description}

\begin{itemize}
\item[\textrm{M1}] (eval, C) tag mutation in the definition of addition;
\item[\textrm{M2}] (eval, Cl) added another clause to the definition of subtraction;
\item[\textrm{M3}] (eval, O) substitution of $-$ for $*$ in arithmetic definitions;
\item[\textrm{M4}] (eval, O) similar to M1 but for conjunction;
\item[\textrm{M5}] (exec, V) bug on assignment; 
\item[\textrm{M6}] (exec, Cl) switch branches in if-then-else;
\item[\textrm{M7}] (exec, Cl) deletion of one of the  \lstinline|while| rule;
\item[\textrm{M8}] (type, C) wrong output type in rule for addition;
\item[\textrm{M9}] (type, C) wrong input type in rule for disjunction.
\end{itemize}





\bibliographystyle{abbrv}

\end{document}